\documentclass[twocolumn]{article}

\usepackage{geometry}
\usepackage{cite}
\usepackage{caption}
\usepackage{graphicx}
\usepackage{amsmath}
\usepackage{amssymb}
\usepackage{textcomp}
\usepackage{authblk}
\usepackage{datetime}
\usepackage{gensymb}
\usepackage{wrapfig}
\usepackage{booktabs}

\newcommand{\onlinecite}[1]{\hspace{-1 ex} \nocite{#1}\citenum{#1}} 

\let\OLDthebibliography\thebibliography
\renewcommand\thebibliography[1]{
  \OLDthebibliography{#1}
  \setlength{\parskip}{0pt}
  \setlength{\itemsep}{0pt plus 0.3ex}
}
  
\title{Does Cosmological Evolution Select for Technology?}
\author[1]{\Large{Jeffrey M. Shainline}
\\
\textit{\large{jeff@soen.systems}}
}
\date{\today}

\begin{document}

\twocolumn[
  \begin{@twocolumnfalse}
    \maketitle
    \begin{abstract}
If the parameters defining the physics of our universe departed from their present values, the observed rich structure and complexity would not be supported. This article considers whether similar fine-tuning of parameters applies to technology. The anthropic principle is one means of explaining the observed values of the parameters. This principle constrains physical theories to allow for our existence, yet the principle does not apply to the existence of technology. Cosmological natural selection has been proposed as an alternative to anthropic reasoning. Within this framework, fine-tuning results from selection of universes capable of prolific reproduction. It was originally proposed that reproduction occurs through singularities resulting from supernovae, and subsequently argued that life may facilitate the production of the singularities that become offspring universes. Here I argue technology is necessary for production of singularities by living beings, and ask whether the physics of our universe has been selected to simultaneously enable stars, intelligent life, and technology capable of creating progeny. Specific technologies appear implausibly equipped to perform tasks necessary for production of singularities, potentially indicating fine-tuning through cosmological natural selection. These technologies include silicon electronics, superconductors, and the cryogenic infrastructure enabled by the thermodynamic properties of liquid helium. Numerical studies are proposed to determine regions of physical parameter space in which the constraints of stars, life, and technology are simultaneously satisfied. If this overlapping parameter range is small, we should be surprised that physics allows technology to exist alongside us. The tests do not call for new astrophysical or cosmological observations. Only computer simulations of well-understood condensed matter systems are required.
    \vspace{3em}
    \end{abstract}
  \end{@twocolumnfalse}
]

	
	
\section{\label{sec:introduction}Introduction}
The ability to devise technology is central to the advancement of society. Certain technologies have become so integral we now depend on them for basic societal operations. Silicon microelectronics is among the most successful technologies in history and has become essential throughout the modern world. Technologies based on superconductivity are also quite enabling. Superconducting tools are important for medical imaging, new forms of information processing, sensitive measurements performed in laboratory science, and large magnets used in particle colliders to study fundamental physics. Most superconducting technologies rely on liquid helium for cooling, and nearly every emerging quantum information platform uses helium to reach temperatures where fragile quantum states can be leveraged. The properties of semiconductors, superconductors, and helium derive from the fundamental physics of the universe, yet fundamental physics and applied technologies are usually studied independently. This paper explores questions at their intersection. How far could the workings of the universe be perturbed and still provide us with transistors? How different could the laws of physics be and still give rise to superconductivity? How improbable is it to find ourselves in a universe in which complex technology can exist at all?

Models of the fundamental physics of the universe require specification of parameters such as coupling strengths and masses of particles. Thirty-one such dimensionless parameters were identified that specify our universe \cite{teag2006}. Fine-tuning refers to the observation that if any of these numbers took a slightly different value, the qualitative features of our universe would change dramatically. Our large, long-lived universe with a hierarchy of complexity from the sub-atomic to the galactic is the result of particular values of these parameters. Physical theories do not offer an explanation of these parameters \cite{sm2004,sm2013,su2003}. The masses and charges of elementary particles are free in the standard model, and many solutions to the equations of string theory appear valid \cite{frsu2004,badi2004,begr2017}. Does similar fine-tuning apply to technology? Would our inventions be sensitive to perturbations of the parameters of nature?

The anthropic principle \cite{bati1986,ba1991,ca1974,care1979,grkr1989,re2000,ba2002,we2006} offers one perspective on fine-tuning: the universe has the parameters it does because they allow life. We could not exist in a universe characterized by significantly different numbers, so we should not be surprised to find ourselves in a universe that enables our existence \cite{bati1986}. The role of life is central to this perspective, leading some to hypothesize the properties of the universe are adjusted for human beings \cite{de1998}. The anthropic perspective does not explain how the universe acquired these parameters. It is unsatisfying to conjecture a universe so peculiar emerged from the vacuum with precisely these improbable specifications. The anthropic principle also does not explain why the universe we inhabit should allow construction of sophisticated technological apparatus. Perhaps any universe conducive to life is also conducive to technology. This article describes means to test this statement.

Smolin introduced the idea that a process of natural selection at the cosmological scale selected the values of the physical parameters specifying our universe \cite{sm1992,sm1994,sm1997,sm2004}, thereby introducing a causal mechanism for the observed fine-tuning. Smolin's theory builds on the established idea that singularities produced by stars in one universe inflate to become offspring universes. A black hole produces a big bang \cite{fagu1986,ha1993,sm1994}. Quantum fluctuations cause the parameters describing the former universe to undergo small mutations before giving rise to the new universe \cite{sm1992,sm1994,wh1973}, introducing a means for evolution through natural selection acting on populations \cite{gaco2013}. Smolin's theory of cosmological natural selection posits an evolutionay trajectory connecting our present universe to an ancestral vacuum fluctuation \cite{sm1997}. This evolutionary process may explain how the parameters of our universe have been selected to maximize the number of offspring produced over the life of the universe through singularities resulting from core-collapse supernovae \cite{sm1992}. The theory may account for the anomalous values of certain physical parameters. The small value of gravitational coupling is necessary for stars to have long lives, while light quark masses lead to nuclear properties necessary for solar fusion. While the anthropic principle leads us to expect physical parameters fine-tuned for life, Smolin's perspective leads us to expect physical parameters fine-tuned for stars. Neither picture accounts for the fitness of the parameters of the universe for the realization of advanced technology. Smolin asked the question ``Why are the laws of physics and the initial conditions of the universe such that stars exist?'' \cite{sm1992} To this I add the question, ``Why are the laws of physics and the initial conditions of the universe such that technology is feasible?'' Perhaps life and technology are both fortuitous consequences of physics tuned for stars, or perhaps they have been selected through the same evolutionary process. Tests described in this paper are designed to differentiate between these possibilities.

Critics of Smolin's hypothesis argue that the parameters do not appear to maximize reproduction through stars \cite{roel1993,el1993,si1997,ha1995} (see also Smolin's counter arguments \cite{sm2004}). Following Smolin's work, others have adopted the perspective that our physical parameters result from an evolutionary process, but conjecture instead that the process has selected for life rather than stars \cite{ha1995,ha1998,ga2003,sm2008,cr2010,st2010,vi2014,pr2017}. Harrison proposed, ``[I]ntelligent life in parent universes creates offspring universes, and in the offspring universes fit for inhabitation, new intelligent life evolves and creates further universes.'' \cite{ha1998} Crane has furthered this perspective with the conjecture that advanced civilizations will eventually create singularities, whether for science or as an energy source \cite{cr2010}. 

For selection to optimize parameters for life, life must have a means to increase the number of offspring. For life as we know it to produce cosmic progeny, technology must be involved. If parameters are tuned for life, they must also be tuned to facilitate technologies necessary for the considerable enterprise of cosmic reproduction. Here I argue Smolin's hypothesis of cosmological natural selection is the process by which our universe acquired its parameters, yet I consider a different outcome related to his statement: ``those choices of parameters that lead to universes that produce the most black holes during their lifetime are selected for.'' \cite{sm1992} Here I ask whether intelligent life equipped with technology could produce more black holes than are produced by stars. An estimate indicates this may be possible, leading to the central hypothesis of this article: Cosmologial natural selection led to parameters co-optimized to enable stars, life, and technology. Stars foster life, life creates technology, and technology accomplishes cosmological reproduction. It is already known the parameters of the universe fall within a narrow range enabling stars and life \cite{ca1974,care1979,bati1986,de1998}, although it is difficult to know exactly how narrow \cite{grkr1989}. If this hypothesis is correct, we should also find the parameters of the universe are tuned to enable specific technologies conducive to reproduction. I conjecture these include semiconductor and superconductor technologies. Silicon has semiconductor properties enabling it to function in an environment conducive to life, while niobium has superconductor properties making it fit to operate in liquid helium. By considering the operation of silicon and the properties of water under the influence of perturbations to the fine-strucutre constant and proton-to-electron mass ratio, we can determine the range of this parameter space across which both silicon and water maintain their fortuitous features. By analyzing niobium alongside helium, we can explore the range of parameters that allows for superconductivity, cooling with liquid helium, and stellar fusion. I propose specific numerical experiments to test the hypothesis.

These numerical experiments may elucidate the relationship between fundamental physics and applied technologies. By conducting these numerical studies, we may learn how broad the range of parameter space is that allows the realization of technologies that are becoming central to the functioning of society and the mission of scientific inquiry. For example, we may find that helium and superconductors retain their desirable properties for any values of physical parameters that give rise to life. In this case, we should not be surprised to find these tools at our disposal. Or we may find that a small sliver of parameter space simultaneously supports the needs of life, helium, and superconductivity. If the numerical tests give this result, we should be surprised to live in a universe with a low-temperate phase diagram of helium that enables cooling of important technologies. Anthropic reasoning does not pertain to this outcome. Cosmological natural selection can account for this result, provided the technologies offered by the universe increase the net fecundity.

This hypothesis is put forth as an extension of Smolin's conception and relies on the same assumptions: new universes can emerge from black hole singularities; and small parameter mutations occur during reproduction. As Smolin states, ``The analogue of biological fitness is then the average number of black holes produced in a universe.'' \cite{sm2013} I further argue that life with technology can produce more black holes than stars alone. An interesting possibility is that there was an era in our evolutionary past when Smolin's picture was complete, and our distant ancestors were optimized for stars. Yet through mutations and selection, the parameters evolved to realize further complexity in life and technology as these adaptations proved useful to produce more offspring.

In Sec.\,\ref{sec:technological_advantage} I describe the physical process by which intelligent technology could produce singularities and argue these means could produce orders of magnitude more offspring than stars alone, pointing to an evolutionary benefit of selecting for technology. Section \ref{sec:technological_coincidences} describes a general approach to identifying instances where the parameters take values that compromise between stars or life and technology. Section \ref{sec:proposed_numerical_studies} proposes specific numerical experiments to test the hypothesis. Discussion of the ramifications is held in Sec.\,\ref{sec:discussion}. 

\section{\label{sec:technological_advantage}Technological Advantage}
For the hypothesis presented here to be correct, technology must provide a means to produce more cosmic offspring than are generated by stellar processes. I assume all offspring produced by supernovae or artificial means result from singularities. Here I use the terms ``singularity'' and ``black hole'' interchangeably, while acknowledging that quantum-gravitational effects may avoid the presence of a true mathematical singularity \cite{asol2018}. This assumption that offspring result from singularities is based on inflationary cosmology \cite{li1982,alst1982,gu1982,gu2004}, which describes the transition of a universe from a very small volume to macroscopic proportions. The theory of inflation is the most empirically successful explanation for the initial conditions that are assumed in standard descriptions of the big bang \cite{gu2004}. The evidence supporting inflation includes the size of the universe as well as its expansion, homogeneity, isotropy, and flatness \cite{gu2004}. Farhi and Guth argued based on the theory of inflation ``that the creation of a universe is necessarily associated with a black hole.'' \cite{fagu1986} Smolin's theory of cosmological natural selection rests on the premise that a black hole in a parent universe is the initial singularity from which a daughter universe inflates \cite{sm1992,sm1994}. As Smolin states, ``[E]ach black hole of our universe leads to such a creation of a new universe and that, correspondingly, the big bang in our past is the result of the formation of a black hole in another universe.'' \cite{sm1994} The new universe is causally disconnected from its parent. I therefore assume here that production of singularities is the means by which technology creates cosmic offspring. If we find it is possible to produce many more black holes technologically than naturally, we must conclude the probability that our universe emerged from a technologically generated singularity in our parent universe is proportionally greater than the probability that we emerged from a singularity resulting from a supernova.

Since the 1980s, the possibility that singularities and subsequent inflated universes could be produced experimentally has been considered \cite{fagu1986,li1991,angu2006}. The task would involve compression of matter (or any energy density) into a sufficiently small volume that it becomes a spacetime singularity in the framework of general relativity. If ordinary matter is used, roughly 10\,kg is required to ensure the singularity inflates \cite{angu2006}. It has been argued that microscopic black holes can be produced by particle colliders \cite{gith2002,kobl2005,la2006}, and measurements at the Large Hadron Collider have been used to put a bound on the minimum black hole mass \cite{cms2011}. However, these small black holes do not have a high probability of inflating into new universes. To lead to new universes, black holes with mass of 10\,kg or larger are required.
 
It has been proposed that advanced civilizations will intentionally produce such black holes in large numbers. Crane argued black holes will be used to serve the needs of the civilizations that create them, with production of offspring a by-product \cite{cr2012,crwe2009}. Tegmark has further argued for the utility of black holes as an energy source \cite{te2017}. When used for power, each source requires one singularity continuing to be fed over time. One singularity produces one offspring, so each power source would produce one daughter universe. This reason for manufacturing singularities does not maximize fecundity. The optimal approach requires intentional and efficient parceling of matter for transformation into singularities. Others have argued the explicit objective of black-hole creation will be to create progeny \cite{ha1995,ha1998,sm2008,st2010,vi2014}. However, it may be the case that the motives will not become clear until far in the technological future. Understanding the motives or evolutionary pressures that lead to creation of offspring is not necessary to formulate and test the hypothesis that our universe has been tuned for the co-existence of stars, life, and technology.

The practical means by which such artificial singularities can be created remains speculative \cite{crwe2009}. Compression may be carried out with high-power lasers or confinement with magnetic fields, two techniques currently being pursued for nuclear fusion. It has been argued that a laser the size of a small asteroid would suffice \cite{cr2010}. Whatever form the apparatus takes, the hypothesis presented here requires that advanced civilizations will be able to accomplish the task of reproduction, aided by the fitness of the universe to realize necessary technologies. In Appendix \ref{apx:technological_advantage} I attempt a comparison between the number of black holes in the history of the Milky Way and the number of progeny that could be realized by converting a single asteroid into singularities. This estimate informs us that $10^9$ core-collapse supernovae have occurred in the history of the Milky Way, while a single, common asteroid could be used to produce $10^{12}$ singularities, indicating the potential for an advanced civilization to produce far more offspring universes than through supernovae alone. It may be possible for a single civilization to convert much of the matter in their solar system to progeny, thereby increasing the fecundity of the universe by orders of magnitude compared to what stars accomplish unaided by intelligence. Even if advanced civilizations are as rare as one per galaxy, life with technology has the potential to create orders of magnitude more offspring than stars.

\section{\label{sec:technological_coincidences}Technological Coincidences}
\begin{figure}[!t]
    \centering{\includegraphics[width=8.6cm]{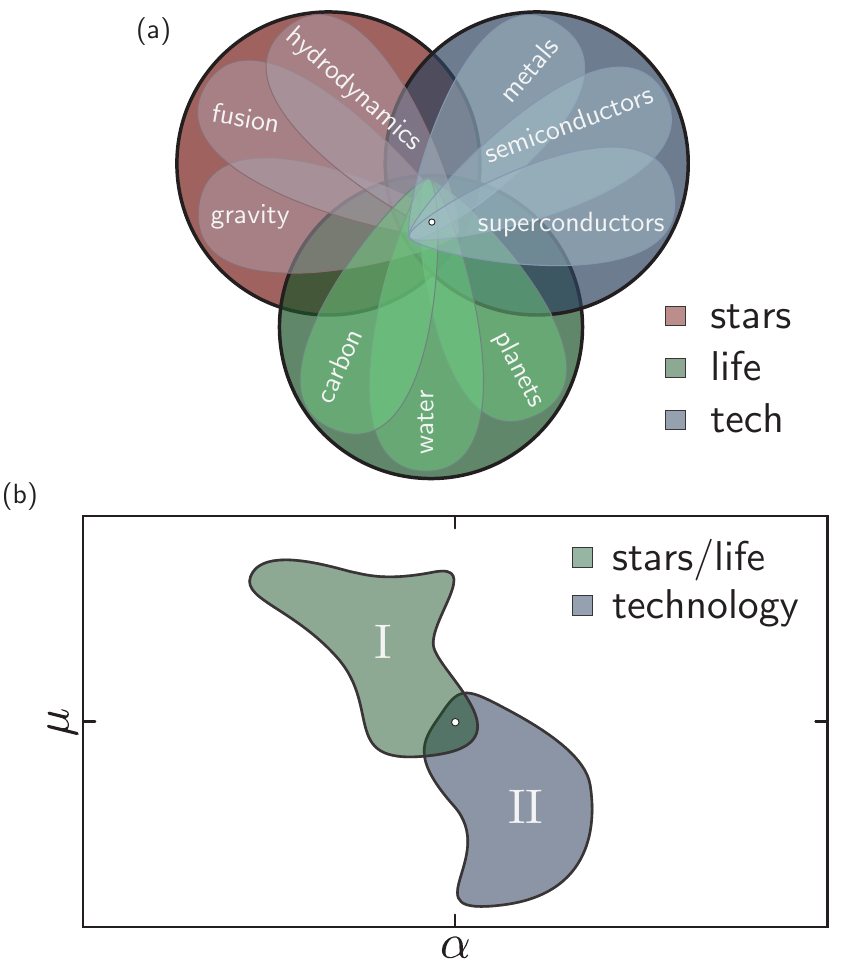}}
	\captionof{figure}{\label{fig:parameter_constraints}Parameter constraints. (a) Venn diagram of the requirements for stars, life, and technology. Relevant sub-regions are identified. The white dot represents the point in parameter space where our universe resides. (b) Schematic of the parameter plane constrained by two considerations. $\mu$ is the proton-to-electron mass ratio, and $\alpha$ is the fine-strucutre constant. The green area may represent the range of values wherein liquid water is more dense than solid, while the blue area may represent the range within which silicon has a band gap and dopant ionization energies useful for digital computing.}
\end{figure}
The goal of the proposed numerical studies is to identify instances in which the constraints placed on physical parameters by technology must be met simultaneously with the constraints of stars or life. This concept is illustrated in Fig.\,\ref{fig:parameter_constraints}. The circles in Fig.\,\ref{fig:parameter_constraints}(a) represent the range of parameters supporting each of these phenomena. Each phenomenon depends on multiple contributions. For example, long-lived stars can only exist if the gravitational coupling constant is within a certain range. Stars also require fusion and hydrodynamical properties that depend on strong nuclear interactions and electromagnetic coupling. Similarly, one may posit that advanced technologies depend on materials including metals, semiconductors, and superconductors. A general survey may identify the range of physical parameters that allow semiconductors with useful band gaps to crystallize as well as the conditions necessary for at least one material to support Cooper-pair formation at finite temperature. Figure \ref{fig:parameter_constraints}(a) illustrates a miniscule subset of relevant considerations for a comprehensive assessment of the range of parameter space enabling stars, life, and technology. Our universe has parameters that fall within this range, depicted by the white dot within the intersecting regions. Many parameter windows have been identified relating to the viability of stars or life \cite{dy1967,da1972,care1979,bati1986,sm1992,sm2004,de1998}, and similar fine-tuning appears to apply to technology.

This general survey indicated by Fig.\,\ref{fig:parameter_constraints}(a) is impractical. Modified parameters may support any of the considered phenomena, but quantifying the effect on fecundity is too complex a task. Further, it may be possible to realize universes bearing little resemblance to our own that are capable of reproduction by different means. Specific inquiries will be more immediately useful in testing the proposed hypothesis. The purpose of this section is to describe several specific inquiries.

Consider the regions of parameter space in which the demands of stars or life overlap those of technology, as shown schematically in Fig.\,\ref{fig:parameter_constraints}(b). If we find through numerical investigation that the parameters of our universe do not fall comfortably within regions suitable for stars or life, but rather are pushed to parameter boundaries by the competing demands of specific technologies, such findings may indicate that cosmological evolution has selected for physical parameters that reach a compromise enabling stars, life, and technology to coexist and maximize fecundity. This search for bounding regions in parameter space is motivated by similar considerations in the context of nuclear physics (for example, see Fig.\,2 of Ref.\,\onlinecite{da1972}).

The two regions of parameter space are labeled I and II, and their areas are $A_{\mathrm{I}}$ and $A_{\mathrm{II}}$. Area is considered illustratively, but the analysis could involve one, two, or more parameters. Co-optimization may be indicated from the areas of the intersection $A_{\cap} = A_{\mathrm{I}}\cap A_{\mathrm{II}}$ (the overlapping region) and the symmetric difference $A_{\bigtriangleup} = A_{\mathrm{I}}\bigtriangleup A_{\mathrm{II}}$ (the sum of the two non-overlapping regions). To assess whether the parameters have been co-optimized under the influence of the two constraints, consider the quantity $c_{T} = A_{\bigtriangleup}/A_{\cap}$. The smallest value $c_{T}$ can take is zero, and it can take arbitrarily large values. If $c_{T}$ is zero, any values of parameters that satisfy one set of constraints also satisfy the other. For example, let $A_{\mathrm{I}}$ represent the region of parameter space in which solar fusion is viable, and let $A_{\mathrm{II}}$ represent the region of parameter space in which superconductivity is viable. If the two regions overlap almost entirely, their symmetric difference vanishes while their intersection remains finite, and $c_{T}\rightarrow 0$. Any choice of parameters that enables stars would also enable superconductivity, and we should not be surprised to find ourselves in a universe hosting superconductors, provided stars are present. Consider instead the scenario in which only small regions of $A_{\mathrm{I}}$ and $A_{\mathrm{II}}$ overlap. In this case, both stars and superconductors are viable over finite regions of parameter space, but only within a much smaller region can they both exist, so $c_{T}$ may be quite large. In this case, we should be surprised to find ourselves in a universe giving rise to both stars and superconductors. I refer to instances of large $c_{T}$ as \textit{technological coincidences}.

To investigate such coincidences, we must conjecture which technologies have been selected through cosmological evolution. At least three classes of technology will be necessary for cosmological reproduction: 1) power sources to fuel the operation; 2) the tools that physically execute the task of forming singularities by compressing energy; and 3) advanced computers to model quantum-gravitational singularities, design the reproduction apparatus, and control the apparatus during operation. This article focuses on the physical implications of the third class. This focus is selected because computers are needed before power sources or black-hole production apparatus, and trends in advanced computing may be indicative of the physical mechanisms required of computer hardware.

The most marked trend in computing is the success of silicon transistors for Boolean logic. Many approaches to logic have been explored with a wide variety of material and physical mechanisms for representing information \cite{ke1985a}. Metal-oxide-semiconductor (MOS) hardware based on silicon has been by far the most fruitful for many reasons. These reasons include: 1) silicon is a semiconductor with a convenient band gap \cite{ke2005}; 2) Si and SiO$_2$ are abundant on rocky planets; 3) Si and SiO$_2$ form a near-perfect interface \cite{grgu2001,wequ2002} to enable metal-oxide-semiconductor field-effect transistors (MOSFETs); 4) the electronic properties of silicon, such as carrier mobility and lifetime, are suitable for implementing high-performance devices \cite{ke1985a}; and 5) due to its chemical and mechanical properties, silicon can be processed with many other materials to realize densely integrated circuits \cite{heza2004}.

In addition to silicon semiconductor technology, a new trend toward superconducting devices for advanced computing is becoming evident. Superconducting circuits can realize qubits for quantum information processing \cite{blga2007,we2017}, neurons for neuromorphic computing \cite{crsc2010,sele2017,scdo2018,sh2019_jap}, and logic gates for digital computing \cite{li2012,hehe2011,hori2013}. Operation at low temperature is required to maintain superconductivity. Niobium is a primary material for superconducting information-processing devices (as well as large magnets) for several reasons: 1) niobium is a superconductor with energy gap significantly larger than the temperature of the liquid-gas phase transition of helium that is used for cooling; 2) niobium is plentiful, perhaps especially where it will ultimately be used \cite{mufo2017}; 3) tunneling barriers can be formed on Nb, enabling Josephson junctions \cite{jo1962,ti1996}; 4) the electronic properties of niobium and related materials enable myriad superconducting devices \cite{vatu1998,ka1999}; and 5) like Si, Nb can be integrated with many other materials for large-scale manufacturing of complex circuits.

More generally, the mechanical, thermal, and electrical properties of metals in conjunction with semiconductors and insulators are unreasonably useful for creation of intricate devices capable of information processing. The functional properties of these three classes of materials are perfectly complimentary for realization of electrical circuitry. If one could make a wish beyond conductors, semiconductors, and insulators, it would be for a class of materials that carry electrical current with no dissipation in a quantum ground state of macroscopic coherence\textemdash a superconductor. We find ourselves in a universe curiously equipped with physics giving rise to a diversity of advanced computational functions based on materials with remarkably useful properties for information processing and perhaps advanced technological intelligence. 

The discussion of the next two sections regarding silicon and niobium is intended to provide specific illustrations of the  general principle that the physics of our universe is tuned not just to make black holes through stars, and not just for life, but also for the creation of advanced technologies. Consideration of other materials may be just as fruitful. Selection for properties of III-V semiconductors and quantum heterostructures may be evident. This idea is not pursued here. The objective of the proposed studies is to find independent bounds on physical parameters within which technology can operate and to compare those bounds to bounds pertinent to stars and life.

\section{\label{sec:proposed_numerical_studies}Proposed Numerical Studies}

\subsection{\label{sec:silicon}Silicon Technology in the \\Context of Life}
For any technology to be useful for large-scale digital computing, the hardware must satisfy several criteria related to materials, devices, and systems \cite{ke1969,ke1982,ke1985a,ke1985b,ke2005}. These criteria include cheap raw materials, insensitivity to device variations and temperature fluctuations, and high gain. Keyes has argued that transistors are unique devices to perform the operations required for digital computing \cite{ke1985a}, and silicon as a material is unique in its ability to yield large numbers of transistors across wafers at low cost. The unique properties of silicon as a material have been detailed by Heywang and Zaininger \cite{heza2004}. Based on their perspective as materials scientists having witnessed the exploration of multiple materials for semiconductor devices, they were led to ask whether the manifestation of silicon in advanced microelectronics is ``a product of man's creativity, custom tailored for his purpose? Or is it\textemdash with its special properties\textemdash still nothing else but a wonderful present of nature?'' \cite{heza2004} Similarly, Keyes has referred to silicon as ``nature's gift to the integrated circuit industry.'' \cite{ke2005} 

If the constants of the universe have evolved to enable silicon technology to coexist with life, it may be informative to simultaneously consider the effect perturbations to fundamental parameters have on the properties of silicon and water. It is often advantageous to formulate such investigations in terms of dimensionless numbers \cite{sabu2018}. For the studies under consideration, the most relevant dimensionless quantities are the proton-to-electron mass ratio, $\mu = m_p/m_e$, and the fine-structure constant, $\alpha = e^2/4\pi\epsilon_0\hbar c$. In the specific case of our universe, these quantities take the values $\mu \approx 1836$ \cite{mu,mone2016}, and $\alpha \approx 1/137$ \cite{mone2016,da2017,payu2018}. The hypothesis predicts the observed values of $\mu$ and $\alpha$ fall in a narrow range that enables water to possess certain properties while also supporting physics conducive to silicon technology. The premise of this argument is that computational technologies based on silicon have augmented the capabilities of intelligent life forms. Because digital computers are capable of solving a wide variety of differential equations modeling many physical phenomena, such general-purpose computers are a tremendous asset for cosmic reproduction. To aid an emerging technological civilization, it is helpful that silicon digital systems operate in the same environment in which life thrives. The temperature of this environment is dictated by the liquid phase of water. 

To grapple with all the features of silicon that make it capable of high-performance digital computers is too large a task. One can reduce the scope of the challenge by focusing on two aspects of the physical system that are important to device operation and are also coupled to the conditions necessary for life. These two aspects are: 1) the energy scales involved, namely the silicon band gap and the ionization energy of dopants in silicon; and 2) the chemical relationship between silicon and oxygen that leads to passivation of surface defects and a robust gate oxide ideal for the construction of a MOSFET. 

Several numerical tests analogous to Fig.\,\ref{fig:parameter_constraints}(b) are possible. In one test, the parameter bounds defining one region may be established based on calculation of the range of parameters that give water specific properties, while the bounds defining the other region may be established from the range of parameters that give silicon a band gap and ionization energy useful for technology. In another test, the parameter bounds suitable for water can be compared to the bounds suitable for a transistor gate oxide. The proposed numerical experiments are summarized in Table \ref{tab:numerical_studies} and further specified in the discussion thereof. Examples of physics that must be tuned to enable silicon microelectronics to coexist with life are now described. Further details of the numerical studies are given in the appendices.

\subsubsection{\label{sec:silicon__energy_gap}The Silicon Band Gap}
For a semiconductor to be useful for computation, there must be a means to alter carrier concentrations. The intrinsic carrier concentration must be much lower than the free-carrier concentration achieved through doping. In a semiconductor, the intrinsic carrier concentration is given by \cite{asme1976}
\begin{equation}
\label{eq:intrinsic_concentration}
n_i \sim (m^*_e m^*_h)^{3/4}(kT)^{3/2}\mathrm{e}^{-E_g/2kT},
\end{equation}
where $n_i$ is the number of valence electrons excited to the conduction band by thermal excitations, $m^*_e$ is the effective mass of an electron, and $m^*_h$ is that of a hole. $E_g$ is the energy of the band gap, and $T$ is the temperature. 

Applying Eq.\,\ref{eq:intrinsic_concentration} to Si, we find $n_i\approx 1.6\times 10^{10}$\,cm$^{-3}$ at $T = 300$\,K. Silicon can be easily doped with phosphorous and boron to create donors and acceptors in excess of $10^{19}$\,cm$^{-3}$. The low intrinsic carrier concentration and high solid solubility of dopants make it possible to create structures with inhomogenous carrier concentrations, a requirement for semiconductor devices such as diodes and transistors. Equation \ref{eq:intrinsic_concentration} depends most strongly on the magnitude of the energy gap through the exponential term. For a semiconductor to be useful for devices, we must have $E_g \gg kT$, and this must be true up to temperatures obtained during operation. In the case of densely integrated silicon transistors clocked at 3\,GHz, this temperature can be close to 400\,K. The operating temperature range of silicon microelectronics is the same as the temperature range of liquid water, although cryogenic operation is also possible \cite{ke1969}.

Another requirement for functional semiconductor devices is the ionization of dopants. For a singly ionized impurity embedded in a crystalline lattice, the ionization energy can be approximated by
\begin{equation}
\label{eq:impurity_ionization_energy}
E_n = \frac{m_e^*}{m_e}\frac{1}{\epsilon_r}\frac{E_{\mathrm{Ry}}}{n^2},
\end{equation}
where $\epsilon_r$ is the relative permittivity of the medium, $E_{\mathrm{Ry}} = 13.6$\,eV is the Rydberg energy, and $n$ is the principal quantum number of the electron to be ionized. One would like the ionization energy to be small enough that all dopants are ionized at 300\,K. In silicon, $\epsilon_r$ is on the order of 10, providing a useful reduction in this ionization energy. The same is true for GaAs. High permittivity leads to screening of the Coulomb interaction between the electrons/holes and the ionic cores of dopants, dramatically reducing the ionization energy, and making it easy for dopants to affect electrical properties at 300\,K.

A third consideration in making useful electronic devices is the power dissipation during operation. In semiconductor circuits that rely on the modification of carrier concentrations for functionality, the energy per operation depends on temperature and also on the semiconductor band gap. To see this, consider the Shockley equation, which gives the forward-voltage current density through a $p-n$ diode \cite{asme1976},
\begin{equation}
\label{eq:pn_current__energy_gap_dependence}
J_f \sim \mathrm{e}^{(eV_f-E_g)/kT}.
\end{equation}
To drive an appreciable current, a forward voltage on the order of $E_g/e$ is required. Nature faces competing influences. If the silicon band gap were smaller, lower voltage and therefore lower energy could be used for switching. However, errors would then occur due to non-negligible concentrations of thermally excited carriers. This is one of the primary reasons that silicon ($E_g = 1.12$\,eV at 300\,K \cite{asme1976}) is used for digital computing rather than germanium ($E_g = 0.67$\,eV at 300\,K \cite{asme1976}), even though carriers in germanium have higher mobility \cite{ke1985a,heza2004}.

The hypothesis of this paper leads to a specific conjecture: the parameters of the universe are tuned to allow silicon to have a band gap that is as small as possible while maintaining a ratio of intrinsic to doped carrier concentrations sufficiently low to enable digital electronic devices to function with low errors. This conjecture can be tested by calculating the silicon band gap and carrier concentrations as a function of $\mu$ and $\alpha$. Given a relationship between intrinsic carrier concentration, doped carrier concentration, and errors in microelectronic circuits \cite{st1977}, modifications to the band gap can be related to device performance.

\subsubsection{The Si/SiO$_2$ Interface as a Transistor Gate Insulator}
Section \ref{sec:silicon__energy_gap} claimed the silicon band gap is near optimal for digital computing at temperatures of liquid water. One may argue that at least one element was bound to be a semiconductor with a decent band gap for this purpose. GaAs has a similar gap. To strengthen the case that Si has been selected through cosmological evolution to enable digital computing, consider also the Si/SiO$_2$ interface. The two most important roles of the surface oxide for transistors are passivation of surface defects and creation of an insulating barrier in the transistor gate \cite{grgu2001}. Numerical investigations of passivating and insulating properties as well as the thermodynamic stability of the Si-SiO$_2$ interface may establish bounds on the values of $\mu$ and $\alpha$ that can support silicon microelectronic technology.

While free carriers in bulk GaAs have significantly higher mobility than in silicon, GaAs has not replaced Si for integrated circuits despite significant investment. The primary reasons for the superiority of Si over GaAs relate to materials, and these properties are determined by fundamental physical parameters. The inability to produce a passivating, insulating oxide on GaAs comparable to SiO$_2$ is the primary limitation \cite{ke1985a}. The Si/SiO$_2$ interface passivates surface defects that would otherwise diminish carrier lifetimes and provides an insulating gate that makes MOS devices possible. Of all the candidate semiconductors on the periodic table, no other material forms a native oxide with properties as desirable as silicon. Even before oxidation, the chemistry between hydrogen and silicon results in a passivated surface with almost no recombination centers \cite{yaal1986}. Considerations pertinent to surfaces may seem less important than electronic properties of bulk materials, but surface defects present on GaAs are often the performance-limiting factor \cite{haal2018}, although material defects in bulk are pervasive and problematic as well. 

To be used as an insulator for a MOSFET gate contact, a dielectric must have a large band gap as well as a high dielectric strength. In these regards, SiO$_2$ is outstanding with a band gap of 9\,eV (nearly ten times that of Si) and a dielectric strength of $10^7$\,V/cm (over 20 times that of air) \cite{wequ2002}. The resistivity of SiO$_2$ is $10^{15}\,\Omega\cdot\mathrm{cm}$, 10 orders of magnitude greater than the resistivity of high-purity Si. For functional circuits, the electrical resistivity of SiO$_2$ must be extremely high, while its thermal resistivity must be low to enable cooling. Indeed, the thermal restivity of SiO$_2$ is only about two orders of magnitude higher than that of Si. 

These properties of large gap and dielectric strength enable a thin layer of SiO$_2$ to function as a gate dielectric in Si transistors, bringing the advantage of operation at low voltage and thus low power. One of the most striking features of silicon microelectronics has been the continued reduction of feature sizes, summarized by Moore's Law \cite{mo1965,calu2012}, enabling denser circuit integration with higher performance at lower cost per device steadily over time. Moore's-Law scaling requires consistent reduction of all relevant feature sizes \cite{ke2005}, including the thickness of the transistor gate insulator. This scaling reaches a limit when the insulator thickness approaches the characteristic length of the electronic wave functions, and tunneling through the barrier becomes non-neligible \cite{kole2013}. The Si/SiO$_2$ interface can be reduced to about five atomic layers while maintaining low leakage current \cite{muso1999,grgu2001,wequ2002}.

In the context of the present study, these properties of SiO$_2$ can be used to formulate another numerical test. By considering the atomic and electronic properties of SiO$_2$ in thin interfacial layers as functions of $\mu$ and $\alpha$, numerical investigation may establish bounds in this parameter plane within which SiO$_2$ is useful as a gate dielectric for Si transistors. These studies can include calculation of the thickness at which the dielectric strength degrades as well as calculation of thermodynamic instability leading to defects. Comprehensive simulation of SiO$_2$ interfaces in conjunction with MOS device operation has already been carried out for the values of physical parameters observed in the present universe \cite{zhli2008}. The hypothesis predicts that the band gap and dielectric strength of SiO$_2$ are as large as possible and that SiO$_2$ insulating layers can be made as thin as possible given other constraints on $\mu$ and $\alpha$ for realizing stars, life, and other advantageous technologies. The observed parameters of our universe allow a gate oxide only a few atoms thick to serve as an excellent insulator in silicon transistors, indicating physical parameters near optimal for enabling high-performance MOSFETs.

\subsubsection{\label{sec:water}The Liquid Phase of Water}
The premise of this section is that silicon microelectronic technologies have been selected through cosmological evolution to enable operation within the temperature range where life resides, which is set by the temperature range of liquid water. Numerical investigations must also explore the effects of changing $\mu$ and $\alpha$ on the properties of water. 

Numerous properties of water affect its ability to foster life \cite{bati1986,de1998}. It is important that water is in the liquid phase across a temperature range conducive to biochemical reactions involving carbon compounds. It is also crucial that the solid phase of water has lower density than the liquid phase so ice will float, leading to high reflection of solar radiation that has been necessary for climate stability throughout the planet's history \cite{bati1986}. Similarly, the optical properties of water vapor and clouds are important for filtering the electromagnetic radiation incident upon the atmosphere. These properties appear fine-tuned to enable functionality at the level of the cell, the organism, and the host planet as a whole. 

The relevant properties of water result from an interplay involving three factors: the intramolecular covalent bonds between the constituent hydrogen and oxygen atoms; the intermolecular hydrogen bonds (H bonds) between a proton in one water molecule and an oxygen atom in another; and van der Waals dispersion forces. In water, the H-O covalent bonds are roughly an order of magnitude stronger than H bonds, while H bonds are an order of magnitude stronger than van der Waals forces. Numerical simulations of water must accurately capture each of these interaction mechanisms spanning several orders of magnitude in strength and spatial scale, a feat accomplished in recent work \cite{disa2014,chko2017,gagu2018}. First-principles calculations of water have correctly calculated the relative densities of the liquid and solid phases \cite{chko2017}, providing the opportunity to investigate these densities as a source of constraints to be used in identification of technological coincidences.

Another instance in which the constraints of life can be juxtaposed those of technology is the carbonate-silicate cycle \cite{waha1981,ka2019}. This feedback loop between the concentration of CO$_2$ in the atmosphere, the temperature of the planet, and the weathering of silicate rocks provides a crucial stabilizing influence that has been shown to extend the habitable zone \cite{kawh1993,kora2013}. Consideration of the carbonate-silicate cycle alongside silicon microelectronics may provide another fertile context to seek technological coincidences, but due to the complexity of the related computations and the requirement for empirical input, we omit this subject from further discussion.

\subsection{\label{sec:niobium}Niobium Technology in the \\Context of Stars}
Section \ref{sec:silicon} argued that silicon technology has been co-optimized alongside life to provide digital computers that operate in the same environment as biological organisms. The present section describes the hypothesis that technologies leveraging superconductivity are also highly advantageous for facilitating cosmic reproduction. No known materials superconduct within the temperature range of liquid water. Even if such materials are discovered, superconducting technologies operating at low temperature are likely to outperform superconducting technologies operating at higher temperature for many applications due to noise and material considerations. The low-temperature superconductors under consideration are well-known materials such as niobium with physics captured by the Bardeen-Cooper-Schreifer (BCS) model of superconductivity \cite{baco1957,ti1996}.

Superconductivity is unreasonably useful in technological contexts, but for cosmological evolution to have selected physical parameters that enable superconductivity, the phenomenon must contribute to greater cosmic fecundity. There are at least two classes of superconducting technologies that may contribute: large magnets used in particle colliders and fusion reactors; and information-processing technologies. If intelligent species are to produce offspring universes, mastery of the physics of our universe is required. Particle colliders are indispensable for this purpose. To carry out the act of cosmic reproduction, tremendous amounts of energy are required. Fusion reactors may be employed in the process or for other purposes within the civilization. Not only can superconductors carry supercurrent with no dissipation, they can carry large supercurrents. One can imagine a universe in which superconductivity exists, but critical current densities are low, making the phenomenon useless for these technologies. Large currents are required to generate magnetic fields that steer particles in colliders and contain plasmas in fusion reactors. If ohmic metals were the only materials available to conduct large currents, these technologies would not be feasible.

In the case of digital computing with silicon, I argued that information-processing technologies are necessary for many functions that facilitate creation of offspring. The same is true of computational systems based on superconductors. One may argue it is redundant and therefore wasteful for nature to provide an excellent platform for digital computing in the form of silicon microelectronics only to see it replaced a few decades later with a digital computing platform based on superconductors. Computational systems based on Josephson junctions \cite{jo1962,ti1996} and other superconducting electronic components \cite{vatu1998,ka1999} appear to be evolving toward functions that augment rather than surmount silicon. Superconducting circuits are promising for quantum information processing \cite{blga2007,we2017} as well as neural computing \cite{crsc2010,sele2017,sh2019_jap,scdo2018}. While superconducting circuits can perform digital logic \cite{li2012,hehe2011,hori2013}, silicon is likely to reign supreme in this domain, possibly into the asymptotic future of technology. Digital superconducting circuits may find the most relevance as a means to interface with and control superconducting qubits or neurons \cite{limc2019}. We may find that technological entities of extraordinary intelligence combining principles of digital logic, quantum computing, and neural information are equipped to produce cosmic progeny. Superconductivity will be of unsurpassed utility in such an entity.

Here again we consider the properties of a specific material: niobium. In the case of Si, we considered the relevant energy scale (the semiconductor band gap) relative to the thermal environment in which it operates (the temperature range of liquid water). In the case of Nb, we consider the relevant energy scale (the superconducting energy gap) relative to the thermal environment in which the superconducting circuits operate (the gas-to-liquid phase transition temperature of liquid helium). Helium plays a significant role in enabling superconducting technologies due to the phase transition at a convenient temperature of 4.2\,K. The superconducting phase transition in niobium occurs at 9.26\,K. Much like errors occur in semiconductor circuits when $kT$ approaches $E_g$, operation of a superconductor at temperature $T$ close to the critical temperature $T_c$ degrades performance.

The utility of superconductors is significantly augmented by the thermodynamic properties of helium for refrigeration, which derive from electronic as well as nuclear degrees of freedom. Yet the nuclear properties of helium are constrained by the requirements for solar fusion. We can therefore conceive of further numerical tests analogous to Fig.\,\ref{fig:parameter_constraints}(b). For example, one bounding region may be determined based on requirements for superconductivity, while another is determined based on the requirements for helium phase transitions. Similarly, a bounding region may be set by the proton-proton fusion chain. These studies may identify technological coincidences wherein the constraints of stars and technology are both improbably satisfied.

\subsubsection{\label{sec:niobium__energy_gap}The Niobium Energy Gap}
\begin{figure}[tb]
    \centering{\includegraphics[width=8.6cm]{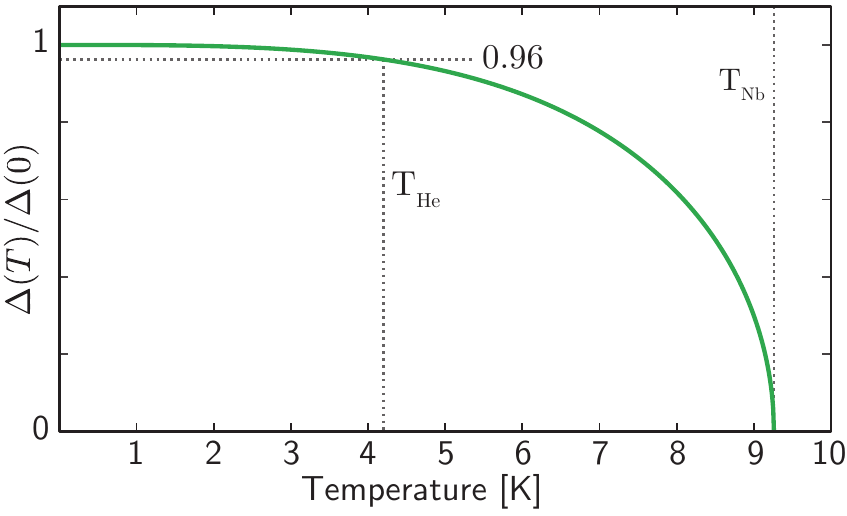}}
	\captionof{figure}{\label{fig:superconducting_gap}Approximate normalized superconducting energy gap of niobium as a function of temperature.}
\end{figure}
For the superconducting technologies mentioned above, one of the most important properties is the superconducting energy gap, $\Delta$. This quantity determines the temperature at which the superconducting state is realized as well as the critical current density that can be carried by the superconductor as a function of temperature. BCS theory provides the relation $T_c \propto \Delta(0)$ for the critical temperature in terms of the gap at zero temperature (see Eq. 3.30 of Ref.\,\onlinecite{baco1957}). The temperature variation of the energy gap does not have a closed-form expression \cite{baco1957}, but it can be approximated as
\begin{equation}
\label{eq:energy_gap}
\frac{\Delta(T)}{\Delta(0)} \sim \bigg[1-\bigg(\frac{T}{T_c}\bigg)^{3.3}\bigg]^{1/2},
\end{equation}
where $\Delta(0)$ is the energy gap at zero temperature and $T_c$ is the superconducting transition temperature (see Ref.\,\onlinecite{ka1999}, pg. 57). Equation \ref{eq:energy_gap} is plotted in Fig.\,\ref{fig:superconducting_gap} for the case of $T_c = 9.26$\,K. The most notable aspect of Fig.\,\ref{fig:superconducting_gap} for the present consideration is that the value of the energy gap is near constant until the temperature of liquid helium, and it drops rapidly above this temperature. When operating at 4.2\,K, the superconducting gap of Nb is 96\% of its value at zero temperature. If the phase transition temperature of helium were 9\,K instead of 4\,K, Nb technology would be far more cumbersome.

In the superconducting domain, Josephson junctions (JJs) are the device of choice for many computational technologies. JJs consist of two superconducting leads separated by a thin tunneling barrier \cite{jo1962,ti1996}. Much like transistors have particular properties that make them uniquely capable of digital computation \cite{ke1985a}, JJs are exceptionally capable of information processing \cite{blga2007,we2017,li2012,hehe2011,hori2013,crsc2010,sh2019_jap} and other technological functions \cite{vatu1998,ka1999}. One important property is the ability of a JJ to produce small, quantized pulses of magnetic flux (fluxons) \cite{ti1996,vatu1998,ka1999} that can represent various types of information. The energy required to produce a fluxon is given by
\begin{equation}
\label{eq:jj_switching_energy}
E_j = I_c\frac{\hbar}{2e},
\end{equation}
where $I_c$ is the critical current of the junction. This critical current depends on the superconducting material used as well as the properties of the tunneling barrier and the area of the junction. One can control $I_c$ across a broad range through lithographic fabrication techniques. The energy of fluxon production depends only on fundamental constants and a parameter, $I_c$, that can be adjusted to suit the application.

For a JJ to be useful, it must not produce fluxons due to random thermal fluctuations, yet it must produce fluxons with sufficiently low energy that large-scale systems can be implemented with power density low enough for cooling with liquid helium. In practice, $I_c \approx 100$\,\textmu A is common in superconducting computational circuits. With this value of $I_c$, $E_j = 3\times 10^{-20}$\,J. Information-processing operations with JJs can be accomplished with a few tens of zeptojoules when operating at $T_{\mathrm{He}}$. This low energy per operation enables dense integrated circuits to operate at high speeds with power density low enough for heat to be removed by helium. Circuits based on Josephson junctions can operate in the 100\,GHz range, another remarkable property that we do not discuss in detail here.

By considering the fine-tuning of parameters enabling superconductivity, we introduce nucleons to the numerical study. In superconductivity, the phonon-mediated attractive interaction depends on the masses of nuclei in the lattice. The most successful first-principles calculations of the niobium phase transition temperature rely on full quantum-mechanical treatment of electronic and nuclear degrees of freedom \cite{luma2005,malu2005}. Considerations pertinent to superconductivity must be addressed in the context of liquid helium, which also derives its properties in part from nuclear interactions.

\subsubsection{\label{sec:helium}Phase Transitions in Helium}
Helium has remarkable features that are unreasonably convenient for cooling to low temperature. No other substance has a gas-to-liquid phase transition anywhere near $T_{\mathrm{He}}$, and no other substance remains liquid at zero temperature. Beyond simply having a single phase transition that is useful for cryogenic applications, bosonic ${}^{4}$He has two useful phase transitions at low temperatures (4.2\,K and 2.2\,K), while fermionic ${}^{3}$He has its own superfluid phase \cite{le1997}. The liquid-gas phase transition at 4.2\,K is commonly employed to cool superconducting computational circuits. This phase transition occurs sufficiently below $T_c$ of important superconductors to enable critical current densities close to the zero-temperature values and for integrated circuits to operate with low errors for circuit parameters giving energy-efficient operation. The liquid-liquid quantum phase transition \cite{mino1975} near 2\,K gives rise to superfluid helium and is used to cool superconducting electromagnets such as those used at the Large Hadron Collider \cite{evbr2008,leta2014}. This transition to the superfluid phase arises from Bose-Einstein condensation, leading to a state that is extraordinary for cryogenic cooling due to its transport properties. These properties include: low viscosity allowing it to penetrate magnet windings; large specific heat ($10^5$ times that of a normal conductor per unit mass); and high thermal conductivity ($10^3$ times that of cryogenic-grade copper) \cite{leta2014}. The phase separation of ${}^3$He and ${}^4$He and the associated heat of mixing enables further cooling into the sub-kelvin regime in dilution refrigerators \cite{uh2012}. Millikelvin temperatures are required to observe quantum coherence in a variety of physical systems that are presently being explored for quantum information processing \cite{we2017,zwdz2013}. The use of helium as a coolant in myriad technological applications has been explored thoroughly \cite{ta2000,leta2014,ro2004,bori2012,toba2010}.

The physical origin of these remarkable properties of helium has been investigated for nearly 80 years \cite{la1941,fe1953,fe1954,fe1955,mino1975,whce1979,cepo1986,seri1986,ce1995,le1997,liri2002,cepa2009,prce2010,prce2017}. The unique physics arises because helium is a quantum fluid with de Broglie wavelength $\lambda_T = \hbar (2\pi/mk_{\mathrm{B}}T)^{1/2}$ on the order of the mean interparticle distance \cite{le1997}, requiring that nuclear exchange and quantum statistics be incorporated \cite{fe1955,le1997}. The phase transitions of ${}^4$He (${}^{3}$He) are rooted in Bose (Fermi) statistics, but the specific values defining the phase diagram and the thermal transport properties depend on parameters of the universe, such as the fine-structure constant and proton-to-electron mass ratio. First-principles numerical investigation of phase transition temperatures in He require two- and three-body interactions be accurately modeled \cite{cepa2009,prce2010,prce2017}. As with superconductors, these interactions depend on both the electron charge and nucleon masses, which enter into calculation of the single-particle wave functions as well as the interaction potentials, again providing a means to incorporate parameters governing both electrons and nucleons in the survey. The numerical investigation proposed in this section relates to the phase-transition temperatures as well as thermal transport properties of helium as a function of $\mu$ and $\alpha$.

Section \ref{sec:silicon} argued silicon is fortuitously equipped to perform digital computations in the temperature range of liquid water to serve as an enabling technology to be discovered by intelligent life. The same argument does not apply to superconductors. Enabled in part by simulations and designs run on silicon computers, superconducting technologies require more sophistication than present-day silicon computers, particularly regarding refrigeration to the temperature of liquid helium. It is certainly possible to achieve large-scale cryogenic systems capable of cooling information processing technologies \cite{ra2009,uh2012} as well as large magnets for colliders \cite{ta2000,evbr2008,ro2004,bori2012} and fusion reactors \cite{toba2010} when operated in Earth's atmosphere. However, in the long term, superconducting technologies are naturally fit for construction and operation outside Earth's atmosphere under cold vacuum conditions. The cosmic microwave background temperature of $2.7$\,K \cite{fi2009} facilitates cooling of superconducting entities that may operate in space. Geological evidence suggests physics related to the interplay between silicon and niobium in the building blocks of the solar system results in present-day asteroids rich in niobium \cite{mufo2017}. Future technologies leveraging superconductivity may reside in space and operate submerged in liquid helium. Whether in a terrestrial environment or in cold space, helium is ideal for cooling superconductors. Consideration of electronic and nuclear constraints related to helium and superconductivity may therefore lead to identification of technological coincidences analogous to silicon and water.

\subsubsection{\label{sec:stars}The Proton-Proton Chain Reaction in Stars}
If cosmological natural selection has arrived at parameters in part to enable helium to cool superconductors, technological coincidences may also be present between helium phase transitions and solar fusion reactions. In solar fusion, the proton-proton ($p\,p$) chain reaction is the dominant means by which stars with mass near that of the sun convert hydrogen to helium \cite{dese2008}. The first step in the $p\,p$ chain involves the nuclear reaction
\begin{equation}
\label{eq:pp_chain__deuterium}
2\,{}^{1}\mathrm{H}\rightarrow {}^{2}\mathrm{H}+e^{+}+\nu_e,
\end{equation}
where ${}^{1}\mathrm{H}$ represents a proton, ${}^{2}\mathrm{H}$ is a deuterium nucleus, $e^{+}$ is a positron, and $\nu_e$ is an electron neutrino. Following production via reaction \ref{eq:pp_chain__deuterium}, the deuterons combine again with protons to form a ${}^{3}$He nucleus via the reaction
\begin{equation}
\label{eq:pp_chain__helium3}
{}^{2}\mathrm{H}+{}^{1}\mathrm{H}\rightarrow {}^{3}\mathrm{He}+\gamma,
\end{equation}
where $\gamma$ is a high-energy photon. Subsequently, in stars of mass near the sun, ${}^{3}$He nuclei accumulate to sufficient levels for ${}^{4}\mathrm{He}$ to be produced via
\begin{equation}
\label{eq:pp_chain__helium4}
2\,{}^{3}\mathrm{He}\rightarrow {}^{4}\mathrm{He}+2\,{}^{1}\mathrm{H}.
\end{equation}
The reactions \ref{eq:pp_chain__deuterium}-\ref{eq:pp_chain__helium4} define the $p\,p$ I branch of the $p\,p$ chain, with branches II and III producing heavier elements. For the present purpose, the objective is to study the nuclear reactions of the $p\,p$ I branch as a function of physical parameters to identify bounds within which the $p\,p$ chain can support a stable solar life cycle, and to compare these bounds to those obtained from phase transitions in He as well as the bounds within which Nb has a $T_c$ well above the ${}^{4}$He liquid-gas phase transition, providing further opportunities to identify technological coincidences.

In the standard solar model, rates of the reactions \ref{eq:pp_chain__deuterium}-\ref{eq:pp_chain__helium4} are used as inputs. These reaction rates are derived from the reaction cross sections, and established means of calculating these cross sections from first principles are summarized in Appendix \ref{apx:stars}. The central challenge for the present purpose is to investigate these nuclear interactions as a function of $\mu$ and $\alpha$ to arrive at modified cross sections. The $p\,p$ chain has been well studied for over eighty years \cite{becr1938,bama1969,kaba1994,scst1998,adau1998,adga2011}, and thus the theoretical infrastructure required to carry out this numerical investigation is established. The mathematical techniques used to assess uncertainties associated with inputs to the model \cite{adga2011} can be used to assess expected deviations in reaction rates when well-known physical parameters ($\mu$, $\alpha$) are intentionally altered, providing a straightforward means to implement the proposed study.

\subsection{\label{sec:numerical_experiments__summary}Summary of Numerical Studies}
\begin{table*}
  \centering
  \begin{tabular}{cccc} 
	\toprule[1.5pt]
	\textbf{Test \#} & \textbf{Systems involved} & \textbf{Bounds 1} & \textbf{Bounds 2} \\ \midrule
	1 & Si/H$_2$O & $n_i(T_{\mathrm{H_2O}})/n(T_{\mathrm{H_2O}})\le \kappa$ & $\rho_{\mathrm{H_2O}}^{s} < \rho_{\mathrm{H_2O}}^{l}$ \\ \midrule
	2 & SiO$_2$/H$_2$O & $I^{\mathrm{leak}}_{\mathrm{SiO_2}}(V_g) \le \kappa$ & $\rho_{\mathrm{H_2O}}^{s} < \rho_{\mathrm{H_2O}}^{l}$ \\ \midrule
	3 & Nb/He & $T_{\mathrm{Nb}}/T_{\mathrm{He}}\ge \kappa$ & $\kappa_1\le T_{\mathrm{He}} \le \kappa_2$ \\ \midrule
	4 & Stars/He & $\tau_{\odot} \ge \kappa$ & $\kappa_1\le T_{\mathrm{He}} \le \kappa_2$ \\
	\bottomrule[1.5pt]
  \end{tabular}
  \caption{Summary of proposed numerical studies.}
  \label{tab:numerical_studies}
\end{table*}
The proposed studies are designed to identify instances wherein the constraints of technology are juxtaposed those of stars or life. In these studies, Si and SiO$_2$ are compared to H$_2$O, while Nb is compared to He. Together, the physics of hydrogen and helium determine much of the behavior of stars. By considering silicon alongside water and niobium alongside helium we may search for a variety of technological coincidences involving only these five elements, yet incorporating diverse physics related to stars, life, and technology. 

The proposed numerical experiments are summarized in Table \ref{tab:numerical_studies}. The experiments fall into two categories: those seeking technological coincidences between silicon technologies and life; and those seeking technological coincidences between superconducting technologies, helium phase transitions, and stars. In test one, the bounds of one region in parameter space are defined by the condition that the ratio of intrinsic carriers in silicon at the temperature of the liquid-solid phase transition of water to the carriers achievable by ionization of dopants [$n_i(T_{\mathrm{H_2O}})/n(T_{\mathrm{H_2O}})$] is less than a threshold determined by consideration of errors in digital logic. Throughout Table \ref{tab:numerical_studies}, $\kappa$ represents any cutoff parameter or stringency criterion that must be chosen. The bounds of the second region are defined by the values of $\mu$ and $\alpha$ that lead to the density of solid water ($\rho_{\mathrm{H_2O}}^{s}$) being lower than the density of liquid water. Calculations can be performed as a function of the relevant parameters, and variation of bounds with respect to $\kappa$ can be analyzed in post processing without the need for additional first-principles calculations. Test two makes use of the same criterion for water, but compares to bounds obtained from the SiO$_2$ transistor gate. As a function of $\mu$ and $\alpha$, the minimum thermodynamically stable oxide thickness that retains the bulk band offset can be calculated. The criterion requires the leakage current through the gate oxide at this thickness is less than a chosen value when a voltage equivalent to the band gap ($V_g = E_g/e$) is applied.

Tests three and four relate properties of helium to stars and superconductors. The bounds related to He may be determined by investigating those values of $\mu$ and $\alpha$ that enable helium to have a liquid-gas phase transition at a temperature ($T_{\mathrm{He}}$) within a range useful for refrigeration. Alternatively, constraints on He may be constructed based on the transition temperature of the superfluid phases of ${}^{3}$He or ${}^{4}$He. In test three, these bounds are compared to bounds obtained by identifying the region of parameter space wherein the ratio of the superconducting phase transition of niobium, $T_{\mathrm{Nb}}$, to that of helium, $T_{\mathrm{He}}$, is greater than a threshold on the order of the value observed in our universe (approximately 2.2). In test four, bounds obtained from consideration of phase transitions in helium can be compared to bounds obtained from fusion in stars by determining the range in parameter space that leads to stars with mass similar to the Sun and lifetime longer than a cutoff required for biological evolution of intelligence followed by evolution of technology and cosmological reproduction. 

In each of these tests, the objective is to determine whether or not technological coincidences, as defined in Sec.\,\ref{sec:technological_coincidences}, are observed in our universe. These studies have been constructed in terms of the proton-to-electron mass ratio, $\mu$, and the fine-structure constant, $\alpha$, because these are the dimensionless parameters most closely related to the physics of the systems under consideration, but construction of similar tests based on other dimensionless parameters could be similarly conceived. The appendices summarize numerical techniques relevant to all these tests, and from cited literature it is apparent these numerical investigations are possible today.

\section{\label{sec:discussion}Discussion}
I argue Smolin's hypothesis is correct: the parameters of the universe evolved through natural selection to maximize fecundity. Smolin proposed stars dominate production of offspring through black holes produced by supernovae. I argue a universe can produce far more offspring if intelligent life uses technology to intentionally convert energy into singularities. This reasoning leads to the hypothesis that the parameters of our universe have been tuned through cosmological evolution to enable stars, life, and technologies conducive to reproduction. We should not be surprised to find ourselves in a universe enabling technology, provided technology makes future, similar universes more likely. The extension to selection for technology does not change the naturalistic physical or philosophical foundations of Smolin's idea. 

While Smolin argued that selection for stars provides a natural time scale for cosmology compatible with current observations of the density parameter ($\Omega$), the same can be said of selection for technology. This natural time scale is the time required for stars to form, give rise to intelligent life, and develop sophisticated technologies. The time scale over which spiral galaxies continue abundant production of stars is the same order of magnitude as the time scale over which technology can be expected to emerge. A key strength of Smolin's work is that ``a hypothesis about particle physics and quantum gravity may be refuted or verified by a combination of astrophysical observation and theory.'' \cite{sm1994} The extension of cosmological evolution to selection for technology provides further benefit in this regard by providing means to test the theory using only calculations of well-understood condensed matter systems. These tests can be performed presently by specialists in each field. If these tests do no reveal technological coincidences, such a finding does not nullify Smolin's original construction of cosmological evolution based on reproduction via stars. Instead, finding that the parameter space that enables stars is similar to the parameter space that enables technology may indicate that once the parameters of the universe have evolved to enable structures as complex as stars, the ability to give rise to technology immediately follows. This finding will provide no information regarding whether technology will eventually be used to produce offspring.

This article has directed attention to several instances in which physics appears tuned for specific technologies. These include the energy scale of the silicon band gap relative to the temperature range of liquid water; the fortuitous nature of the Si/SiO$_2$ interface; and the energy scale of the niobium superconducting gap relative to phase transitions in helium. The hypothesis predicts the parameters of the universe take values within a narrow range enabling the technologically useful properties of these materials, and that slightly adjusted parameters that improve semiconductor or superconductor device performance can be selected only at the expense of stars, life, or other technologies. Technological coincidences may also be investigated without consideration of stars or life by comparing the domains of validity of two technologies. For example, SiO$_2$ is important in transistors and also for fiber-optic communication. The optical absorption of SiO$_2$ across the wavelength range of established light sources can be analyzed as a function of $\mu$ and $\alpha$ for comparison with the functional range of MOSFETs identified by test two in Table \ref{tab:numerical_studies}. Further tests can be conceived, perhaps with other materials such as compound semiconductors or with carbon in place of water.

Independent of cosmological ramifications, it may be valuable to understand the fundamental physics underpinning the technologies that enable future progress in society. Helium is particularly interesting in this regard, as it is central to cooling many systems of growing technological significance, while the physics providing the useful thermodynamic properties are rooted in fundamental statistical mechanics. The critical temperature of the liquid-to-gas phase transition of ${}^{4}$He has been briefly considered for differing values of the mass of the atom \cite{seri1986}. Further investigation of the low-temperature phase diagram as a function of fundamental physical parameters may provide valuable insights into the physics that enable the technologies on which society depends.

The reasoning presented here differs from anthropic reasoning as described by Carter: ``[W]hat we can expect to observe must be restricted by the conditions necessary for our presence\ldots.'' \cite{ca1974,we2006} This observer bias does not apply to technology. One can imagine a universe allowing intelligent life that did not allow transistors. If we find technological coincidences, they cannot be explained by anthropic principles, but they may be evidence of selection for technology in cosmological evolution.

While departing from anthropic reasoning, a central ramification of the hypothesis still relates to the role of intelligent life \cite{st2010}. In the proposed model, we are not the sole purpose of existence, nor are we an irrelevant fluke. Along with stars and technology, we may be part of an interacting system that develops within the universe along a physically guided trajectory, culminating in large numbers of progeny. We should not be surprised that the conditions of the universe allow our existence, if we play a role in the life cycle. We should also not be surprised that technologies of tremendous sophistication are feasible with materials that are ubiquitous in the solar system and across the galaxy, produced by stars, if these technologies are important in the evolutionary process. As we proceed to invent/discover new technologies, we may expect to find tools suited to the long-term survival of civilization culminating in production of singularities. We should expect to discover the technologies nature intends us to utilize.

\section{Acknowledgements}
I appreciate valuable feedback on this manuscript from Dr. Kristan Corwin, Dr. Charles Clark, Dr. Richard Mirin, and Dr. Krister Shalm as well as insights on numerical investigation of the properties of helium from Dr. Tom Bourne, Dr. David Ceperley, Dr. Allan Harvey, Dr. Eric Shirley, Dr. Ray Mountain, and Dr. Michael Moldover. I am endlessly grateful for the support of my wife, Katie Miles Shainline.


\appendix

\section{\label{apx:technological_advantage}Estimate of Technological Advantage}
Black hole singularities occur naturally as the outcome of core-collapse supernovae \cite{be1990,bu2013}. For the hypothesis presented here to be valid, it must be possible for artificial creation of offspring to outpace the natural creation through supernovae. This appendix presents an estimate indicating this technological advantage may be possible.

The present-day rate of core-collapse supernovae in the Milky Way is $r_0=0.019\pm0.0011$/year, as measured by $\gamma$ ray detection of ${}^{26}\mathrm{Al}$ formed in core-collapse supernovae \cite{diha2006}. Observation of the density of gas, dust, metals, and stars in 80 galaxies \cite{wola1995} as a function of redshift \cite{fa1996} as well as observation of the supernova relic neutrino background \cite{kast2000} indicate that the present-day supernova rate is lower than in the past, although the rate as a function of time remains uncertain (see Sec. 18.1 of Ref.\,\onlinecite{dese2008}). Making the assumption that the supernova rate tracks the metal enrichment rate, the maximum rate of core-collapse supernovae was an order of magnitude larger than the present rate \cite{fa1996}. Because the supernova rate as a function of time is difficult to ascertain exactly, we proceed as if it has been constant since the formation of the galaxy at 10 times the present value. This approximation results in a high-end estimate for the number of astrophysical singularities and therefore a conservative estimate for the fitness of technology relative to supernovae for reproduction. We take the lifetime of the Milky Way to be ten billion years, leading to the estimate that $2\times10^{9}$ offspring universes have been created by core-collapse supernovae to date in our galaxy.

One source of energy density that could be used for artificial creation of singularities is rocky matter, such as asteroids, distributed in solar systems. Asteroid sizes are power-law distributed over a broad range, with some as small as a few meters, and the largest in our solar system nearly 100 kilometers in diameter \cite{hada2015,deal2015}. There are $990\pm20$ near-Earth asteroids larger than 1\,km in diameter \cite{hada2015}, and these asteroids are likely to be composed of silicates, iron, and other metals. The density of silicon is 2.3\,g/cm$^3$, and that of iron is 7.9\,g/cm$^3$. Seeking a conservative estimate, we assume the density of an asteroid is 2\,g/cm$^3$. The mass of a 1-km-diameter asteroid is greater than $8.4\times 10^{12}$\,kg, indicating that $10^{12}$ offspring could, in principle, be manufactured from this matter. The energy source used to derive force to compress the matter is likely to be the fusion reactor of the Sun with its output of $4\times 10^{26}$\,W \cite{gr2018}.

\section{\label{apx:silicon}Numerical Studies of Silicon}
Accurate first-principles calculations of the silicon band gap have been conducted since 1985 \cite{huch1985,hylo1985}. The most successful first-principles approach is based on many-body perturbation theory (the so-called GW approximation). This approach was used in Ref.\,\onlinecite{hylo1985}, and the calculated band gap was within 3.4\% of the measured value. More recently, density functional theory (DFT) \cite{hoko1964,kosh1965} has become the most common approach for numerically efficient calculations of properties of solids. Historically, DFT struggled to accurately calculate band gap energies as well as properties of localized defects. Hybrid functionals that incorporate a fraction of non-local Hartree-Fock exchange \cite{peer1996a,hesc2003,pama2006} have been demonstrated to overcome this shortcoming \cite{gasc2016}. Regarding impurities, treatment including electrostatic boundary conditions gives accurate calculations of many common point defects in silicon \cite{sc2006}. In particular, DFT has recently been used to accurately calculate the energy levels of phosphorous donors in silicon \cite{smbu2017}. With these numerical techniques, it is possible to calculate the silicon band gap and dopant energy levels of silicon. By calculating these properties as a function of $e$ and $m_e$ in conjunction with a model for errors in digital computation \cite{st1977} one can investigate the range of parameter space wherein silicon is a functional material for digital computing in the temperature range defined by the liquid phase of water.

\section{\label{apx:silicon_dioxide}Numerical Studies of SiO$_2$}
First-principles calculations of surface oxides have been carried out for several decades based on Hartree-Fock and generalized valence bond methods \cite{rego1981}. The perturbative GW theory can be applied to SiO$_2$ as well \cite{shri2008}. The GW approach has been employed to compute band offsets that determine tunneling and leakage properties. Such calculations have been applied to the Si/SiO$_2$ interface \cite{shri2008} as well as the Si/Si$_3$N$_4$ interface \cite{phli2013}, leading to close agreement with experiment without fitting parameters. One study of particular relevance to the present work combined a number of computational techniques, including first-principles molecular dynamics to identify the atomic structures at the Si/SiO$_2$ interface during thermal oxidation, DFT to obtain band gap profiles, and the non-equilibrium Green's function method to directly calculate leakage current through the gate \cite{kiki2013}. Such an approach represents a promising strategy to holistically analyze the Si/SiO$_2$ interface and MOSFET device performance as a function of $e$ and $m_e$. This work could be extended to include a more accurate means of calculating band offsets, such as GW \cite{shri2008} or DFT with empirical fitting \cite{albr2008}. Combined analysis of interface formation with DFT has demonstrated agreement with experimental band offsets as well as insensitivity to specific interface configuration \cite{zhph2017}. These simulation techniques enable accurate calculation of Si/SiO$_2$ surface properties during and after oxide growth as well as during chemical processing steps that are important for device fabrication \cite{trra1990}.



\section{\label{apx:water}Numerical Studies of Water}
Numerical investigations of water have a long history \cite{aupi1968,gial2016} and have struggled to achieve high accuracy \cite{gial2016,weno2004}. Density functional formulations have become the most popular for simulating water. As is the case with many DFT calculations, the primary challenge resides in constructing an appropriate exchange-correlation functional. Numerous such functionals have been proposed and utilized \cite{gial2016}. It was shown in Ref.\,\onlinecite{disa2014} that using a hierarchy of exchange-correlation functionals including both exact exchange and dispersion corrections gives quantitative agreement with measurements of properties such as electronic distribution functions and bond angular distributions in liquid water \cite{disa2014}. An alternative formulation based on the recently introduced strongly constrained and appropriately normed semilocal density functional \cite{suru2015} has achieved high accuracy in matching X-ray diffraction data of electron distribution functions, the structure of the H-bond network, and the density of ice I$_{\mathrm{h}}$ relative to liquid water \cite{chko2017}, a quantity emphasized in the numerical studies proposed here. Dielectric-dependent hybrid functionals have been shown to accurately predict properties of water at temperatures up to 400\,K \cite{gagu2018}. 

\section{\label{apx:niobium}Numerical Studies of Niobium}
The proposed investigations of Nb relate to the critical temperature of the superconducting phase transition, $T_c$. The superconducting transition temperature is known from BCS theory \cite{baco1957} to be related to the superconducting energy gap through the relation $T_c \propto \Delta_0$, where $\Delta_0$ is the superconducting energy gap at zero temperature. We can therefore calculate the transition temperature if we can calculate the superconducting energy gap. Density functional theory has proven quite accurate for this calculation, provided electrons and nucleons are both handled appropriately \cite{luma2005,malu2005}. References \onlinecite{luma2005} and \onlinecite{malu2005} have calculated $T_c = 9.5$\,K, a 2\% error relative to the measured value of 9.3\,K \cite{ei1954,asme1976}. The numerical approach developed in Ref.\,\onlinecite{luma2005} extends the established Kohn-Sham formalism \cite{kosh1965} to treat electrons and nuclei quantum mechanically on the same footing. Such an approach is necessary to accurately model superconductivity, as electron-phonon interactions are central to the pairing mechanism that results in the superconducting ground state. The framework of DFT is adequate for the proposed study that seeks to calculate $\Delta_0$ from first principles as a function of $e$, $m_e$, and $m_p$.

\section{\label{apx:helium}Numerical Studies of Helium}
Most contemporary work on calculations of helium leverage a virial expansion wherein a given property of the system (i.e., the pressure) is expressed as a power series of the density \cite{ma2008}. The viral equation of state is more useful for evaluating transport properties on the system away from phase boundaries. Within the virial equation of state formalism, accurate calculation of the inter-atomic potentials is required. Significant progress has been made in recent years to determine the form of the two-body \cite{ujvi2005,ujvi2006,prce2010,prce2017} and three-body \cite{ujvi2007,cepa2009} interaction potentials, leading to ab initio models of sufficient accuracy to be used as standards to calibrate apparatus used for measurement of transport properties \cite{azja1995,humo2000,gamo2011}.
These interaction potentials are also applicable to the condensed phase of interest in the present study. The variational density-matrix approach has been applied at non-zero temperature to calculate a liquid-gas phase transition in helium within 1\,K of the measured value \cite{seri1986}. Reference \onlinecite{seri1986} explored the effect of modifying the helium mass on the liquid-gas phase transition temperature, pointing to the disappearance of the liquid-gas phase transition in helium with mass less than 10\% its observed value, a phenomenon predicted to hold in general systems obeying Bose statistics independent of the specific interactions \cite{mino1975}. This disappearance of the liquid phase of interacting bosons at zero temperature for sufficiently small mass introduces a lower limit on the mass of He if its liquid phase is to be used to cool superconducting technologies. 

The path integral formalism may be the most accurate approach to first-principles calculations of the properties of condensed helium near phase transitions \cite{cepo1986,ce1995}. This formalism allows calculations of the many-body density matrix to arbitrary accuracy, in principle. In practice, the Monte-Carlo path-integral approach is accurate enough that Ceperley and Pollock were able to calculate the correct value of the superfluid phase transition temperature in ${}^{4}$He with the computational resources available in 1986 \cite{cepo1986}. The fitness of silicon microelectronics for digital computation has ensured far greater computational resources are available today, leading to the potential for even higher accuracy and fast caculation across a range of parameter values. Incorporation of three-body and more accurate two-body interactions is likely to further improve this technique for calculation of phase transitions from free energies.

\section{\label{apx:stars}Numerical Studies of Fusion}
In the standard solar model, rates of the reactions \ref{eq:pp_chain__deuterium}-\ref{eq:pp_chain__helium4} are used as inputs. For reaction \ref{eq:pp_chain__deuterium}, the rate is too slow to be measured in a laboratory, so direct calculation from weak-interaction theory is the only means to attain this rate. For this reaction, potential methods have been quite successful. This reaction was first treated in Ref.\,\onlinecite{becr1938}, and theoretical treatments have since been refined with more accurate wave functions and a more realistic nuclear transition operator \cite{scst1998}. The theoretical treatment given in Ref.\,\onlinecite{scst1998} has compared five nucleon-nucleon interaction potentials and found excellent agreement between all five models and available experimental data such that the dominant source of uncertainty results not from the choice of potentials but from uncertainty in the measured coupling strengths. In the present context, these wavefunctions and matrix elements must be calculated as a function of the relevant fundamental physical parameters to obtain new scattering cross sections, while the same measured coupling constants can be employed. Effective field theory methods may provide an alternative means to attain similar accuracy \cite{pama2003} in calculating the cross section of reaction \ref{eq:pp_chain__deuterium}. Reference \onlinecite{pama2003} demonstrated the use of an effective field theory of quantum chromodynamics in a formalism enabling parameter-free prediction of nuclear cross sections that agree with potential methods. Reaction \ref{eq:pp_chain__helium3} can also be treated with potential methods \cite{casc1998,mavi2005}. The approach of Ref.\,\onlinecite{mavi2005} provides accuracy within 1\% of that extrapolated from measurements at higher energy. Reaction \ref{eq:pp_chain__helium4} is the most directly constrained by experiments of all reactions in the $p\,p$ chain. Due to the availability of direct measurements, the cross section of this reaction is well known without first-principles modeling. Nevertheless, theoretical treatments have been carried out. Reference\,\,\,\onlinecite{tybl1991} treated this reaction using the semi-phenomenological resonating group method, providing reasonable agreement with measured data. Similar potential methods or effective field theory calculations as described above may be capable of providing more accurate theoretical treatments of reaction \ref{eq:pp_chain__helium4}, but the availability of high-quality measured data has obviated the need for such analysis.
	
\bibliographystyle{unsrt}	
\bibliography{selection_for_technology}

\end{document}